\documentclass[prd,twocolumn,superscriptaddress,altaffilletter,nofootinbib]{revtex4}
\usepackage[dvips]{graphicx}
\usepackage{amsmath}
\usepackage{graphicx,epsfig}
\newcommand{\be}{\begin{equation}}
\newcommand{\ee}{\end{equation}}
\newcommand{\bea}{\begin{eqnarray}}
\newcommand{\eea}{\end{eqnarray}}
\newcommand{\der}{\partial}

\begin{document}


\title{Global asymptotic dynamics of Cosmological Einsteinian Cubic Gravity}


\author{Israel Quiros}\email{iquiros@fisica.ugto.mx}\affiliation{Dpto. Ingenier\'ia Civil, Divisi\'on de Ingenier\'ia, Universidad de Guanajuato, Gto., M\'exico.}

\author{Ricardo Garc\'ia-Salcedo}\email{rigarcias@ipn.mx}\affiliation{CICATA-Legaria, Instituto Polit\'ecnico Nacional, Ciudad de México, CP 11500, México.}

\author{Tame Gonzalez}\email{tamegc72@gmail.com}\affiliation{Dpto. Ingenier\'ia Civil, Divisi\'on de Ingenier\'ia, Universidad de Guanajuato, Gto., M\'exico.}

\author{Jorge Luis Morales Mart\'inez}\email{jorge.morales@cimat.mx}\affiliation{Divisi\'on de Ingenier\'ia, Universidad de Guanajuato, Gto., M\'exico.}

\author{Ulises Nucamendi}\email{unucamendi@gmail.com}\affiliation{Instituto de F\'isica y Matem\'aticas, Universidad Michoacana de San Nicol\'as de Hidalgo, Edificio C-3, Ciudad Universitaria, CP. 58040 Morelia, Michoac\'an, M\'exico.}\affiliation{Departamento de F\'isica, Cinvestav, Avenida Instituto Polit\'ecnico Nacional 2508, San Pedro Zacatenco, 07360, Gustavo A. Madero, Ciudad de M\'exico, M\'exico.}

\date{\today}


\begin{abstract} In this paper we investigate the cosmological dynamics of an up to cubic curvature correction to General Relativity (GR) known as Cosmological Einsteinian Cubic Gravity (CECG), whose vacuum spectrum consists of the graviton exclusively and its cosmology is well-posed as an initial value problem. We are able to uncover the global asymptotic structure of the phase space of this theory. It is revealed that an inflationary matter-dominated bigbang is the global past attractor which means that inflation is the starting point of any physically meaningful cosmic history. Given that higher order curvature corrections to GR are assumed to influence the cosmological dynamics at early times -- high energies/large curvature limit -- the late-time inflation can not be a consequence of the up to cubic order curvature modifications. We confirm this assumption by showing that late-time acceleration of the expansion in the CECG model is possible only if add a cosmological constant term.\end{abstract}


\maketitle


\section{Introduction}\label{sec-intro}

Higher curvature corrections to general relativity have become popular at present time when we are looking for a ``compass'' pointing to a right direction where to find answers to the many unsolved puzzles of contemporary physics. Higher-order corrections to GR are required by the renormalization procedure to work \cite{stelle, stelle-1, book-1, book-2, hindawi, hindawi-1}. Generalizations of general relativity are considered as gravitational alternatives for unified description of the early-time inflation with late-time cosmic acceleration in \cite{nojiri-rev-2011}. Among the modified theories considered are, the $F(R)$ and Horava-Lifshitz $F(R)$ gravity, scalar-tensor theory, string-inspired and Gauss-Bonnet theory, non-local gravity, non-minimally coupled models, and power-counting renormalizable covariant gravity. It was shown in that reference that some versions of the mentioned theories may be consistent with local tests and may provide qualitatively reasonable unified description of inflation with a dark energy epoch. 


The higher curvature modifications of GR are characterized by the high complexity of their mathematical structure. In this case only through given approximations one may retrieve some useful analytic information on the cosmological dynamics. Otherwise one has to perform either a numeric investigation or one have to the apply the tools of the dynamical systems theory. By means of the dynamical systems tools one obtains very useful information on the asymptotic dynamics of the mentioned cosmological models. The asymptotic dynamics is characterized by: i) attractor solutions to which the system evolves for a wide range of initial conditions, ii) saddle equilibrium configurations that attract the phase space orbits in one direction but repel them in another direction, iii) source critical points which may be pictured as past attractors, or iv) limit cicles, among others.

Although the use of the dynamical systems is specially useful when one deals with scalar-field cosmological models -- see references \cite{ellis-book, coley-book, wands-prd-1998, faraoni-grg-2013, bohmer-rev, quiros-rev, quiros_ejp_rev} for a very small but representative sample of related research -- their usefulness in other contexts has been explored as well \cite{quiros1, quiros2, quiros3, quiros4}. In \cite{quiros1} by means of a combined use of the type Ia supernovae and $H(z)$ data tests, together with the study of the asymptotic properties in the equivalent phase space the authors demonstrated that the bulk viscous matter-dominated scenario is not a good model to explain the accepted cosmological paradigm. Meanwhile, in \cite{quiros2} the authors explored the whole phase space of the so called Veneziano/QCD ghost dark energy models where the dynamics of the inner trapping horizon is ignored, and also the more realistic models where the time-dependence of the horizon was taken into consideration. In a similar way, in \cite{quiros3} it was investigated to which extent noncommutativity -- a property of quantum nature -- may influence the cosmological dynamics at late times/large scales. It was enough to explore the asymptotic properties of the corresponding cosmological model in the phase space. The dynamical systems tools were also applied in \cite{quiros4} to study the asymptotic properties of a cosmological model based on a non-linear modification of General Relativity in which the standard Einstein-Hilbert action is replaced by one of Dirac-Born-Infeld type containing higher-order curvature terms.


In a recent paper \cite{arciniega-plb-2020} an up to cubic curvature correction to GR was proposed, with the following features: (i) its vacuum spectrum consists of a transverse massless graviton exclusively, just as in GR, (ii) it possesses well-behaved black hole solutions which coincide with those of Einstein cubic gravity (ECG) \cite{bueno-prd-2016, bueno-prd-2016-1, chinese}, (iii) its cosmology is well-posed as an initial value problem and (iv) it entails a geometric mechanism triggering an inflationary period in the early universe (driven by radiation) with a graceful exit to a late-time cosmology arbitrarily close to $\Lambda$CDM. 

In the present paper we shall look for the global asymptotic dynamics of the CECG model proposed in \cite{arciniega-plb-2020}. Our aim is to correlate the generic solutions of the model with past and future attractors as well as with saddle equilibrium configurations in some state space. This will give a solid mathematical basis to several statements made in \cite{arciniega-plb-2020}. It will be confirmed, in particular, that non-standard matter-dominated inflationary Friedmann evolution is the global past attractor of any phase space orbits that represent viable cosmic histories. Unlike this, our results will show that the statement that graceful exit to a late-time $\Lambda$CDM cosmology in the CECG model is a consequence of the proposed curvature modification of GR, is incorrect. As a matter of fact it will be shown that late-time de Sitter expansion in the CECG scenario is possible only if add a cosmological constant term. A similar study of the so called $f(P)$ cubic gravity have been presented in \cite{marciu}.


We have organized the paper in the following way. In section \ref{sect-setup} we expose the basic elements of the CECG model, including the cosmological equations of motion. Then, in section \ref{sect-ds} we trade the 2nd order cosmological field equations by a set of autonomous ordinary differential equations (ODE) on some phase space variables, which we identify with the dynamical system of the model. In this section we find the critical points of the resulting dynamical system and study their existence and stability properties. In order to illustrate our study with numeric computations a phase portrait of the model is drawn. The particular case of the CECG model without the cosmological constant is explored in section \ref{sect-no-lambda} in order to elucidate the role of the vacuum energy in the global asymptotic dynamics. In section \ref{sect-discuss} we discuss on the most important physical aspects resulting from the dynamical systems investigation and in section \ref{sect-conclu} brief conclusions are given. Finally, in order for the paper to be self-contained we have included an appendix section \ref{sect-app} where we give very simplified exposition of the fundamentals of the dynamical systems theory which are useful in most cosmological applications.

\bigskip


\section{The CECG model}\label{sect-setup}

In \cite{arciniega-plb-2020} a cubic modification of Einstein's GR was proposed which generalizes the so called Einsteinian cubic gravity (ECG) \cite{bueno-prd-2016, bueno-prd-2016-1, chinese}. The proposed modification rests on the following combination of cubic invariants: ${\cal P}-8{\cal C}$, where

\begin{widetext}\bea &&{\cal P}=12R_{\mu\;\;\lambda}^{\;\;\nu\;\;\sigma}R_{\nu\;\;\sigma}^{\;\;\tau\;\;\rho}R_{\tau\;\;\rho}^{\;\;\mu\;\;\lambda}+R_{\mu\lambda}^{\;\;\;\;\nu\sigma}R_{\nu\sigma}^{\;\;\;\;\tau\rho}R_{\tau\rho}^{\;\;\;\;\mu\lambda}-12R_{\mu\lambda\nu\sigma}R^{\mu\nu}R^{\lambda\sigma}+8R_\mu^{\;\;\lambda}R_\lambda^{\;\;\nu}R_\nu^{\;\;\mu},\nonumber\\
&&{\cal C}=R_{\mu\lambda\nu\sigma}R^{\mu\lambda\nu}_{\;\;\;\;\;\;\tau}R^{\sigma\tau}-\frac{1}{4}\,RR_{\mu\lambda\nu\sigma}R^{\mu\lambda\nu\sigma}-2R_{\mu\lambda\nu\sigma}R^{\mu\nu}R^{\lambda\sigma}+\frac{1}{2}\,RR_{\mu\nu}R^{\mu\nu}.\label{cubic-inv}\eea\end{widetext} The CECG is based on the following action:\footnote{In \cite{oliva} it shown that the combination of cubic invariants defining five-dimensional quasitopological gravity, when written in four dimensions, reduce to the CECG. It is also introduced a quartic version of the CECG and a combination of quintic invariants with the properties of the mentioned theory. Meanwhile in \cite{arciniega-2} it is shown how to construct invariants up to 8th order in the curvature.}

\bea S=\frac{1}{2}\int d^4x\sqrt{|g|}\left[R-2\Lambda+2\beta\left({\cal P}-8{\cal C}\right)+2{\cal L}_m\right],\label{action}\eea where $\Lambda$ is the (non-negative) cosmological constant and $\beta$ is a non-negative free parameter and ${\cal L}_m={\cal L}_m(g_{\mu\nu},\psi)$ is the Lagrangian of the matter degrees of freedom $\psi$. Usually, besides the two polarizations of the graviton, there may exist two massive modes: a ghosty graviton with mass $m_g$ and a scalar mode with mass $m_s$. In the present theory the massive modes decouple: $m_g\rightarrow\infty$, $m_s\rightarrow\infty$.

In terms of Friedmann-Robertson-Walker line-element (flat spatial sections): $ds^2=-dt^2+a^2(t)\delta_{ik}dx^idx^k$, the cosmological equations of motion derived from \eqref{action} read \cite{arciniega-plb-2020}:

\bea &&3H^2\left(1+16\beta H^4\right)=\rho_m+\Lambda,\nonumber\\
&&2\dot H\left(1+48\beta H^4\right)=-(p_m+\rho_m),\label{moteq}\eea together with the continuity equation $\dot\rho_m=-3H(\rho_m+p_m)$. In what follows, for simplicity, we assume the following equation of state for the matter fluid: $p_m=\omega_m\rho_m$, where the constant $\omega_m$ is the equation of state (EOS) parameter.


\begin{figure*}[t!]
\includegraphics[width=7cm]{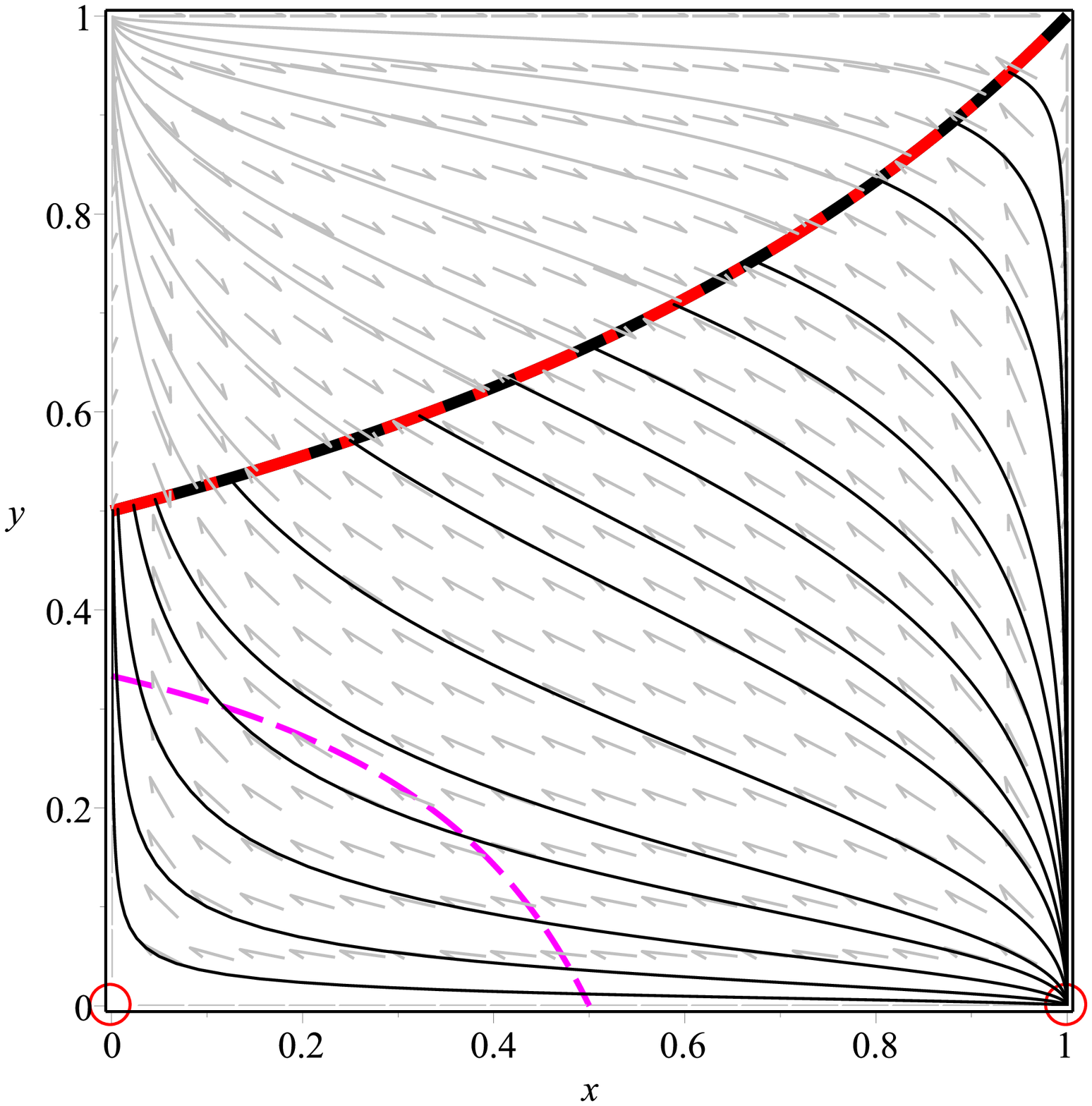}
\includegraphics[width=7cm]{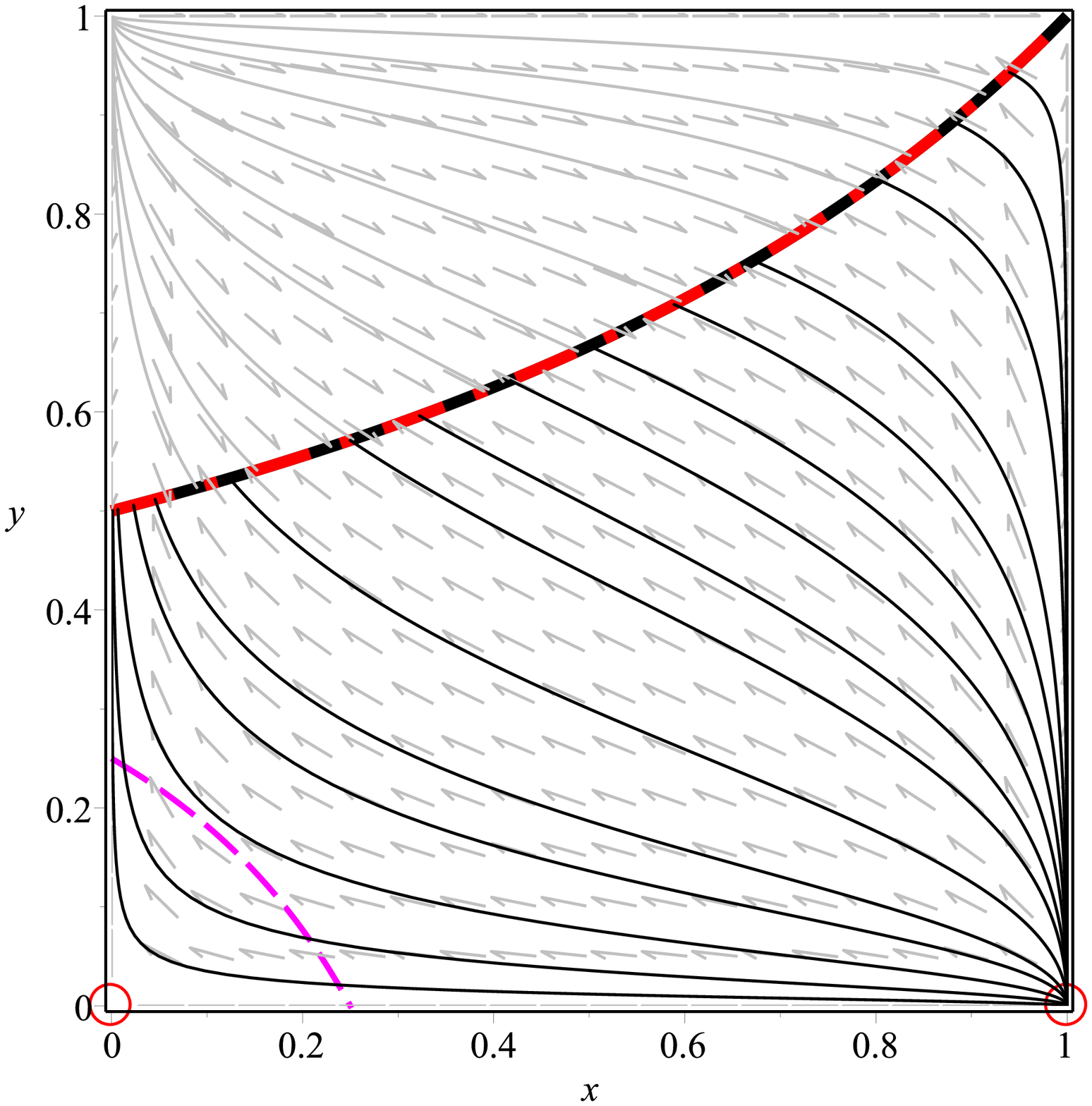}
\vspace{1.5cm}\caption{Phase portrait of the dynamical system \eqref{ode} for the radiation, $\omega_m=1/3$ (left-hand panel) and for dust, $\omega_m=0$ (right-hand panel). The critical points of the dynamical system appear enclosed by the small circles, while the de Sitter attractor manifold ${\cal M}_\text{dS}$ \eqref{b-psi}, is represented by the dash-dot curve that joints the points $(0,1/2)$ and $(1,1)$. The physically meaningful region of the phase space -- properly $\Psi$ in \eqref{psi} -- lies below this curve which coincides with the upper boundary $\der\Psi$ (black solid curve). The gray orbits do not entail physically meaningful cosmic evolution. The dashed curve around the origin in the bottom-left corner of the drawings (curve given by \eqref{ac-dc}) enclose the regions where decelerated expansion occurs.}\label{fig1}\end{figure*}



\section{Dynamical System}\label{sect-ds}

We introduce the following bounded variables of some phase space:

\bea x\equiv\frac{16\beta H^4}{1+16\beta H^4},\,y\equiv\frac{\Lambda}{3H^2+\Lambda},\label{vars}\eea where $0\leq x\leq 1$ and $0\leq y\leq 1$. The modified Friedmann constraint -- first equation in \eqref{moteq} -- can be written in the following way:

\bea \Omega_m\equiv\frac{\rho_m}{3H^3}=\frac{1-(2-x)y}{(1-x)(1-y)},\label{fried-const}\eea meanwhile,

\bea &&\frac{\dot H}{H^2}=-\frac{3(\omega_m+1)(1-x)}{2(1+2x)}\,\Omega_m\nonumber\\
&&\;\;\;\;\;\;\;=-\frac{3(\omega_m+1)\left[1-(2-x)y\right]}{2(1+2x)(1-y)}.\label{hdot}\eea 

In terms of the phase space variables $x$, $y$, the second-order cosmological equations \eqref{moteq} may be traded by the following two-dimensional autonomous dynamical system:

\bea &&\frac{dx}{dv}=\frac{6(\omega_m+1)x(1-x)\left[(2-x)y-1\right]}{1+2x},\nonumber\\
&&\frac{dy}{dv}=-\frac{3(\omega_m+1)y(1-y)\left[(2-x)y-1\right]}{1+2x},\label{ode}\eea where we have introduced the time variable $v=\int(1+\Lambda/3H^2)Hdt$.

The phase space where to look for equilibrium configurations of the dynamical system \eqref{ode} is the following region of the $(x,y)$-plane:

\bea \Psi=\left\{\left(x,y\right):\,0\leq x\leq 1,\;0\leq y\leq\frac{1}{2-x}\right\}.\label{psi}\eea The boundary

\bea \der\Psi=\left\{\left(x,y\right):\,0\leq x\leq 1,\,y=\frac{1}{2-x}\right\},\label{b-psi}\eea separates the the physically meaningful region of the phase space where $\Omega_m\geq 0$, from the unphysical region where $\Omega_m<0$. 

Another curve of physical interest is the one related with the change of sign of the deceleration parameter:

\bea q\equiv-1-\frac{\dot H}{H^2},\label{q}\eea i. e., the curve that follows from the condition $q=0$,

\bea y=\frac{3(\omega_m+1)-2(1+2x)}{3(\omega_m+1)(2-x)-2(1+2x)}.\label{ac-dc}\eea


\subsection{Critical points and their properties}

The critical points $P_i:(x_i,y_i)$ of the dynamical system \eqref{ode} in the phase space $\Psi$, as well as their stability properties, are listed and briefly discussed below.

\begin{enumerate}

\item Inflationary bigbang solution, $P^\text{infl}_\text{bb}:(1,0)$. The eigenvalues of the linearization matrix evaluated at this point are: $$\lambda_1=\omega_m+1,\;\lambda_2=2(\omega_m+1).$$ Hence, this is the source point (global past attractor) and is characterized by: $$x=1\Rightarrow H^4\gg\frac{1}{16\beta},\,y=0\Rightarrow 3H^2\gg\Lambda,$$ which leads to the following modified Friedmann equation: 

\bea 48\beta H^6=\rho_m.\label{p1-fried}\eea In this case $\Omega_m$ is undefined while, $$\frac{\dot H}{H^2}=-\frac{1}{2}(\omega_m+1),$$ so that for the deceleration parameter \eqref{q} we get: $q=(\omega_m-1)/2$. Since for physically meaningful matter $0\leq\omega_m\leq 1$ $\Rightarrow-1/2\leq q\leq 0$, this means that the critical point $P^\text{infl}_\text{bb}$ is to be associated with accelerated expansion. This is why we call it as ``inflationary bigbang'' to differentiate it from standard bigbang.

\item Matter domination, $P_\text{mat}:(0,0)$. Given that the eigenvalues of the linearization matrix at $P_\text{mat}$: $$\lambda_1=-6(\omega_m+1),\;\lambda_2=3(\omega_m+1),$$ are of different sign, this means that the matter-dominated solution is a saddle critical point. At this solution $\Omega_m=1$ $\Rightarrow$ $3H^2=\rho_m$, and $$\frac{\dot H}{H^2}=-\frac{3}{2}(\omega_m+1)\Rightarrow q=\frac{3\omega_m+1}{2}.$$

\item de Sitter attractor manifold: $${\cal M}_\text{dS}:\left(x,\frac{1}{2-x}\right),\;0\leq x\leq 1.$$ For points in ${\cal M}_\text{dS}$ we obtain the following eigenvalues of the corresponding linearization matrix: $$\lambda_1=0,\;\lambda_2=3(\omega_m+1)\left(\frac{x-1}{2-x}\right).$$ The vanishing eigenvalue is associated with an eigenvector that is tangent to the manifold at each point. The second eigenvalue is always a non-positive quantity. This means that, as seen from the FIG. \ref{fig1}, each one of the critical points in ${\cal M}_\text{dS}$ is a local attractor, i. e., the manifold itself is a global attractor of orbits in $\Psi$. For each point in the de Sitter attractor manifold, $\dot H=0$, $\Omega_m=0$ $\Rightarrow q=-1$.

\end{enumerate} Notice that all of the three critical points above always exist. 

In FIG. \ref{fig1} the phase portrait of the dynamical system \eqref{ode} is shown. The critical points $P^\text{inf}_\text{bb}$ and $P_\text{mat}$ appear enclosed by the small circles, while the de Sitter attractor ${\cal M}_\text{dS}$ is represented by the dash-dot curve which coincides with the upper boundary $\der\Psi$ of the physically meaningful phase space (black solid curve). The gray orbits that are above the boundary do not entail any physically meaningful cosmic evolution pattern. The thick dashed curve encloses the region of the phase space $\Psi$ (bottom-left corner of the phase portrait) where the expansion of the universe is decelerated. Hence, given that decelerated expansion is required for the formation of the amount of observed cosmic structure to happen, only those orbits that go across the enclosed region represent viable cosmic histories. These orbits cross the dashed curve -- equation \eqref{ac-dc} -- twice and the corresponding cosmic histories show two periods of accelerated expansion separated by a period of decelerated expansion when the cosmic structure forms.


\section{CECG model without the cosmological constant}\label{sect-no-lambda}

Let us investigate the role of the cosmological constant in the global asymptotic dynamics of the CECG model. For this purpose we shall study the model with a mix of two fluids: dust and radiation, and without the cosmological constant. In this case the cosmological equations of the CECG model read:

\bea &&3H^2\left(1+16\beta H^4\right)=\rho_d+\rho_r,\nonumber\\
&&2\dot H\left(1+48\beta H^4\right)=-\rho_d-\frac{4}{3}\rho_r,\nonumber\\
&&\dot\rho_d=-3H\rho_d,\;\dot\rho_r=-4H\rho_r,\label{moteq-1}\eea where $\rho_d$ and $\rho_r$ represent the energy density of the dust and of the radiation, respectively. 

We shall trade the above system of 2nd order differential equations by a two-dimensional dynamical system. For this purpose we choose the phase space coordinate $x$ defined in \eqref{vars} and the new $y$-coordinate:

\bea y=\frac{\Omega_r}{1+\Omega_r},\;\Omega_r\equiv\frac{\rho_r}{3H^2}.\label{y-var}\eea Then equations \eqref{moteq-1} are equivalent to the dynamical system:

\bea &&\frac{dx}{dv}=-\frac{2x(1-x)\left[3-(2+x)y\right]}{1+2x},\nonumber\\
&&\frac{dy}{dv}=-2y(1-y)\left[2-2y-\frac{3-(2+x)y}{2(1+2x)}\right],\label{ode-1}\eea where we have introduced the following time variable $v=\int(1+\rho_r/3H^2)Hdt$, and in order to eliminate the $\Omega_d$ terms, we have used the following relationship:

\bea \Omega_d=\frac{1}{1-x}-\frac{y}{1-y}.\label{omega-d}\eea Other useful equations are:

\bea \frac{\dot H}{H^2}=\frac{(x+2)y-3}{2(1+2x)(1-y)},\label{hdot-1}\eea and the equation that follows from requiring that the deceleration parameter vanishes:

\bea y=\frac{4x-1}{3x}.\label{q-1}\eea This curve separates the region in the phase where the expansion is accelerated from the region where it is decelerated.


\begin{figure}[t!]
\includegraphics[width=7cm]{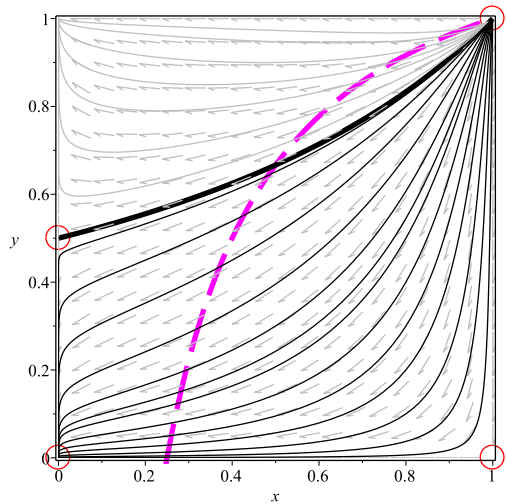}
\vspace{1.5cm}\caption{Phase portrait of the dynamical system \eqref{ode-1}. The critical points of the dynamical system appear enclosed by the small circles. The physically meaningful phase space is the region below the black solid curve (gray orbits do not represent physically meaningful cosmological evolution). The dashed curve encloses the region of the phase space where the expansion is accelerated.}\label{fig2}\end{figure}



\subsection{Equilibrium states}

The critical points $P_i:(x_i,y_i)$ of the dynamical system \eqref{ode-1} are found in the physically meaningful phase space ($\Omega_d\geq 0$):

\bea \Psi=\left\{\left(x,y\right):\,0\leq x\leq 1,\;0\leq y\leq\frac{1}{2-x}\right\}.\label{psi-1}\eea Here we list the existing critical points and briefly comment on their properties, including stability.

\begin{enumerate}

\item Inflationary radiation-dominated bigbang, $P^\text{infl}_\text{rad}:(1,1)$. This solution is the global past attractor to which every orbit of the phase space converges into the past (this is confirmed numerically). In this case the cosmic dynamics is governed by a modified Friedmann equation: $$48\beta H^6=\rho_r\Rightarrow a(t)=\left(\frac{2}{3}\right)^\frac{3}{2}\left(\frac{M_r}{48\beta}\right)^\frac{1}{4}\,t^{3/2},$$ where we have taken into account that $\rho_r=M_r a^{-4}$ ($M_r$ is a constant parameter) and we have arbitrarily set to zero the integration constant.

\item Radiation-dominated Friedmann expansion solution, $P_\text{rad}:(0,1/2)$. For this case the eigenvalues of the linearization matrix are: $\lambda_1=1/2$ and $\lambda_2=-4$, so that this is a saddle equilibrium point. We have that the cosmic dynamics is governed by the standard Friedmann equation: $3H^2=\rho_r$.

\item Inflationary non-standard dust-dominated solution, $P^\text{infl}_\text{dust}:(1,0)$. It is a saddle critical point since the eigenvalues of the corresponding linearization matrix: $\lambda_1=2$ and $\lambda=-3$, are of opposite sign. According to this solution the cosmic dynamics is governed by the modified Friedmann equation: $$48\beta H^6=\rho_d\Rightarrow a(t)=\frac{1}{4}\left(\frac{M_d}{48\beta}\right)^\frac{1}{3}\,t^2,$$ where we have taken into account that $\rho_d=M_d a^{-3}$.

\item Dust-dominated solution, $P_\text{dust}:(0,0)$. The eigenvalues of the linearization matrix: $\lambda_1=-1$, $\lambda_2=-6$, are both negative quantities so that the dust-dominated solution: $3H^2=\rho_d$, is the global attractor.

\end{enumerate}

In FIG. \ref{fig2} a phase portrait of the dynamical system \eqref{ode-1} is drawn. The critical points are enclosed in the small circles. The black solid curve divides the phase space into a physically meaningful region (below the curve) -- properly the relevant phase space $\Psi$ \eqref{psi-1} -- and a region where the orbits do not represent physically meaningful cosmic histories. The dashed curve encloses the region where the expansion is accelerating.

As seen there are two types of cosmic evolution. To the first type belong those orbits that after emerging from the inflationary radiation-dominated past attractor go close enough to the, also inflationary, non-standard dust-dominated solution to finally end up at the global attractor: the standard decelerated expansion matter-dominated solution. The second type consists of orbits that, after emerging from the global past attractor, go close enough to the standard decelerated expansion radiation-dominated solution and then are attracted by the standard matter-dominated solution (the global future attractor). The first type of orbits leads to not as well motivated kind of cosmic evolution as that in the second type since there is not a period of standard radiation-dominated decelerated expansion.


\begin{figure}[t!]
\includegraphics[width=7cm]{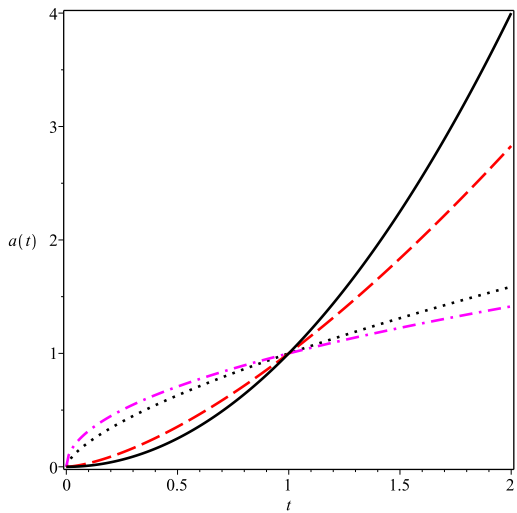}
\vspace{1.5cm}\caption{Types of cosmic evolution: (1) non-standard (inflationary) Friedmann radiation-dominated evolution -- dash, (2) non-standard (inflationary) dust-dominated expansion -- solid, (3) standard (decelerated) radiation-dominated expansion -- dash-dot, and (4) standard dust-dominated evolution -- dots.}\label{fig3}\end{figure}



\begin{figure}[t!]
\includegraphics[width=7cm]{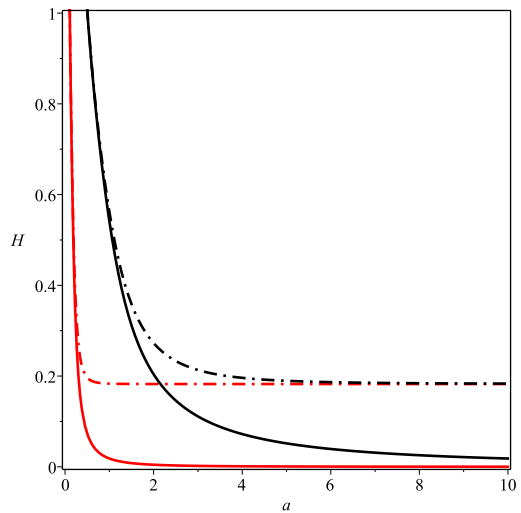}
\vspace{1.5cm}\caption{The drawing of the Hubble rate $H$ vs the scale factor $a$ -- according to \eqref{h-solve} -- is shown for the cases when the cosmological constant $\Lambda$ vanishes (solid curves) and when it is a non-vanishing quantity (dash-doted curves). We consider only expanding cosmology so that in \eqref{h-solve} we take the positive sign. The darker curves are for dust ($\rho_d\propto a^{-3}$), while the remaining ones are for radiation ($\rho_r\propto a^{-4}$). It is seen that for vanishing $\Lambda=0$, as the expansion proceeds the Hubble rate asymptotically vanishes, which means that the end point of the expansion in this case is the static universe.}\label{fig4}\end{figure}



\section{Discussion}\label{sect-discuss}

One of the most interesting consequences of the present scenario is the existence of a matter-dominated inflationary bigbang -- critical point $P^\text{infl}_\text{bb}:(1,0)$ -- which is the global past attractor. Hence, all of the orbits emerge from this unstable equilibrium inflationary state. Given the modified Friedmann equation \eqref{p1-fried} and that $\rho_m=M\,a^{-3(\omega_m+1)}$, where $M$ is a constant, it follows that the scale factor evolves with the cosmic time in the following fashion:

\bea a(t)=\left[\left(\frac{\omega_m+1}{2}\right)^2\left(\frac{M}{48\beta}\right)^\frac{1}{3}\right]^\frac{1}{\omega_m+1}\,t^\frac{2}{\omega_m+1},\label{p1-a}\eea where we have arbitrarily set to zero the integration constant. From \eqref{p1-a} it follows that,

\bea \frac{\ddot a}{a}=\frac{2(1-\omega_m)}{(1+\omega_m)^2}\,t^{-2},\;H(t)=\frac{2}{\omega_m+1}\,t^{-1}.\label{p1-ddota}\eea 

Our results confirm in rigorous mathematical terms the conclusion of \cite{arciniega-plb-2020} that in the CECG scenario primordial inflation is a natural stage from which any plausible cosmic history -- depicted by given orbits in the phase space -- starts. This is to be expected since higher-curvature corrections, such as the cubic ones, are expected to modify the dynamics at early times, i. e., at very high energy/curvature. However, the conclusion in the mentioned reference that it is possible in this scenario to obtain not only primordial inflation but also late time acceleration in purely geometric terms, happens to be wrong. 

In section \ref{sect-no-lambda} we have studied the CECG model with a mix of two fluids: radiation and dust, and with vanishing cosmological constant $\Lambda=0$. It is confirmed that, as stated in \cite{arciniega-plb-2020}, the inflationary radiation-dominated stage driven by non-standard Friedmann equation, $$48\beta H^6=\rho_r,$$ is the global past attractor, i. e., it is the starting point of any orbit in the phase space. However, the global future attractor is the standard dust-dominated decelerated expansion driven by the Friedmann equation $3H^2=\rho_d$. This means that the late-time de Sitter solution in the CECG model with non-vanishing cosmological constant is due, precisely, to the non-vanishing $\Lambda\neq 0$, and is not a curvature effect as stated in \cite{arciniega-plb-2020}.

Let us to elucidate which was the loophole in the analysis in \cite{arciniega-plb-2020} that led to the incorrect conclusion. Their analysis was based on the following equation (equation (9) of \cite{arciniega-plb-2020}):

\bea \frac{\ddot a}{a}=H^2-\frac{(\omega_m+1)\rho_m}{2(1+48\beta H^4)},\label{9}\eea which is obtained by combining $\ddot a/a=H^2+\dot H$ with equations \eqref{moteq}. From this equation it apparently follows that there are two asymptotic stages where $\ddot a/a\approx H^2$ is a positive quantity, so that the expansion occurs at an accelerated pace. This is true at early times, whenever $H^4\gg 1/48\beta$ and $H^4\gg\rho_m/\beta$, as well as at late times as long as the matter density dilutes with the expansion $\rho_m\rightarrow 0$. These conclusions are unjustified since the equation \eqref{9} is misleading. Actually, if substitute $H^2$ in \eqref{9} from \eqref{moteq}, one gets:

\bea \frac{\ddot a}{a}=\frac{\rho_m+\Lambda}{3\left(1+16\beta H^4\right)}-\frac{(\omega_m+1)\rho_m}{2(1+48\beta H^4)}.\label{true-9}\eea It is this equation, and not \eqref{9}, the one that leads to correct analysis. It is seen from \eqref{true-9} that at early times/high curvature, when $H^4\gg 1/48\beta$, $$\frac{\ddot a}{a}\approx\frac{(1-\omega_m)\rho_m+2\Lambda}{96\beta H^4},$$ so that $\ddot a/a\geq 0$ and the expansion is accelerated. At late times, when the density of matter has diluted enough with the curse of the cosmic expansion: $\rho_m\propto a^{-3(\omega_m+1)}$, i. e., in the limit $\rho_m\rightarrow 0$, from \eqref{true-9} it follows that: $$\frac{\ddot a}{a}\rightarrow\frac{\Lambda}{3(1+16\beta H^4)},$$ so that the expansion is accelerated only for non-vanishing $\Lambda>0$. The same conclusion is obtained if solve the algebraic Friedmann equation -- first equation in \eqref{moteq} -- in terms of the Hubble rate:

\bea H=\pm\sqrt\frac{\left(2\sqrt{3\beta}\,\alpha+\sqrt{1+12\beta\alpha^2}\right)^{2/3}-1}{4\sqrt{3\beta}\left(2\sqrt{3\beta}\,\alpha+\sqrt{1+12\beta\alpha^2}\right)^{1/3}},\label{h-solve}\eea where for simplicity of writing we have introduced the notation: $\alpha\equiv\rho_m+\Lambda$. It is seen from \eqref{h-solve} that at early times, i. e., in the formal limit when $\rho_m\rightarrow\infty$ $\Rightarrow\alpha\gg 1/\sqrt\beta$, we get that $$H=\pm\frac{\alpha^{1/6}}{(4\sqrt{3\beta})^{1/3}}\Rightarrow 48\beta H^6=\alpha,$$ as it should be. Meanwhile, at late times, i. e., in the formal limit $\rho_m\rightarrow 0$ $\Rightarrow\alpha=\Lambda$, the Hubble rate is non-vanishing only if $\Lambda\neq 0$. Actually, if take $\Lambda=0$, i. e., $\alpha=0$ in equation \eqref{h-solve}, it is obtained that $H=0$ (static universe). This is illustrated in FIG. \ref{fig4} where the drawing of the Hubble rate $H$ vs the scale factor $a$, according to \eqref{h-solve}, is shown for the cases when the cosmological constant vanishes and when it is a non-vanishing quantity.

Summarizing: at late times accelerated expansion in the CECG model is possible only if add a non-vanishing cosmological constant. Late-time accelerated expansion is not, in any way, a consequence of the higher curvature modifications of the theory.


\section{Conclusion}\label{sect-conclu}

In this paper we have put on solid mathematical grounds the result of previous works \cite{arciniega-plb-2020, arciniega-2} that primordial inflation is the natural starting point of any plausible cosmic history within the framework of the CECG scenario. We have done this on the basis of the dynamical systems analysis of the CECG model. Dynamical systems offer a unique robust information on the generic solutions of the cosmological equations of motion, i. e., those that are preferred by the differential equations according to their structural stability properties.

In the same rigorous manner we have shown that the late-time accelerated de Sitter expansion in the CECG model is a result of considering a non-vanishing cosmological constant and is not related in any way to the effects of the higher curvature contribution $\propto{\cal P}-8{\cal C}$. Our result is natural in the sense that the higher curvature modifications of GR are supposed to have impact in the high-energy, large curvature regime exclusively, i. e., at early times in the cosmic evolution.


\section{Acknowledgments}

We thank Julio Oliva, Mihai Marciu and Shoulong Li, for interesting comments and for pointing us to relevant bibliographic references. We are also grateful to SNI-CONACyT for continuous support of their research activity. The work of RGS was partially supported by SIP20200666, COFAA-IPN, and EDI-IPN grants. UN also acknowledges PRODEP-SEP and CIC-UMSNH for financial support of his contribution to the present research.



\appendix



\section{Remarks on the dynamical systems}\label{sect-app}

Here, for simplicity of the exposition, we consider a 2D phase space. The interplay between a cosmological model and the corresponding phase space is possible due to an existing one-to-one correspondence between exact solutions of the cosmological field equations and points in the phase space spanned by given variables $x$, $y$: $$x=x(H,\rho_i),\;y=y(H,\rho_i).$$ When we replace the original field variables $H$, $\rho_i$, by the phase space variables $x$, $y$, we have to keep in mind that, at the same time, we trade the original set of non-linear second order differential equations in respect to the cosmological time $t$ by a set of first order ODE-s: 

\bea x'=f(x,y),\;y'=g(x,y),\label{asode}\eea where the tilde denotes derivative with respect to certain dimensionless ``time'' parameter $\tau$. 

The most important feature of the system of ODE \eqref{asode} is that the functions $f(x,y)$ and $g(x,y)$ do not depend explicitly on the time parameter. This is why \eqref{asode} is called as an autonomous system of ODE. The image of the integral curves of \eqref{asode} in the phase space are called ``orbits'' of the dynamical system. 

The critical points of the dynamical system \eqref{asode} $P_i:(x_i;y_i)$, i. e., the roots of the system of algebraic equations $$f(x,y)=0,\;g(x,y)=0,$$ correspond to privileged or generic solutions of the original system of cosmological equations. In order to judge about their stability properties it is necessary first to linearize \eqref{asode} around the hyperbolic equilibrium points, where by an hyperbolic critical point $P_h$ it is understood that the real parts of all of the eigenvalues of the linearization matrix around $P_h$ are necessarily non-vanishing. In particular, an hyperbolic point can not be a center. In more technical words: an hyperbolic critical point is a fixed point that does not have any center manifolds.

Linearising around a given hyperbolic equilibrium point $P_i$ amounts to consider small linear perturbations $$x\rightarrow x_i+\delta x(\tau),\;y\rightarrow y_i+\delta y(\tau).$$ These perturbations would obey the following system of coupled ODE (here we use matrix notation):

\bea \delta{\bf x}'=J(P_i)\cdot\delta{\bf x},\;\delta{\bf x}=\begin{pmatrix}\delta x\\\delta y\end{pmatrix},\;J=\begin{pmatrix}\frac{\der f}{\der x}&\frac{\der f}{\der y}\\\frac{\der g}{\der x}&\frac{\der g}{\der y}\end{pmatrix}.\label{matrix-eq}\eea where $J$ - the linearization or Jacobian matrix, is to be evaluated at $P_i$. Thanks to the Hartman-Grobman theorem \cite{hartman, grobman}, which basically states that the behavior of a dynamical system in the neighborhood of each hyperbolic equilibrium point is qualitatively the same as the behavior of its linearization, we can safely replace the study of the dynamics of \eqref{asode} by the corresponding study of its linearization \eqref{matrix-eq}.

We assume that $J$ can be diagonalized, i. e., $J_D=M^{-1} J M$, where $M$ is the diagonalization matrix and $$J_D=\begin{pmatrix}\lambda_1&0\\0&\lambda_2\end{pmatrix},$$ is the diagonal matrix whose non vanishing components are the eigenvalues of the Jacobian matrix $J$:

\bea \det|J-\lambda U|=0,\;U=\begin{pmatrix}1&0\\0&1\end{pmatrix}.\label{caract-ec}\eea

After diagonalization the coupled system of ODE \eqref{matrix-eq} gets decoupled:

\bea \delta\bar{\bf x}'={\it J_D}\cdot\delta{\bar{\bf x}},\;\delta{\bar{\bf x}}={\it M}^{-1}\delta{\bf x},\label{decop-edo}\eea where we have to recall that to each equilibrium point it corresponds a different matrix $J_D$. The decoupled system of ODE \eqref{decop-edo} is easily integrated: $$\delta\bar x(\tau)=\delta\bar x(0)\,e^{\lambda_1\tau},\;\delta\bar y(\tau)=\delta\bar y(0)\,e^{\lambda_2\tau}.$$ Since the diagonal perturbations $\delta\bar x$ and $\delta\bar y$ are linear combinations of the perturbations $\delta x$, $\delta y$: $$\delta\bar x=c_{11}\delta x+c_{12}\delta y,\;\delta\bar y=c_{21}\delta x+c_{22}\delta y,$$ where the constants $c_{ij}$ are the components of the matrix $M^{-1}$, then  

\bea &&\delta x(\tau)=\bar c_{11}\,e^{\lambda_1\tau}+\bar c_{12}\,e^{\lambda_2\tau},\nonumber\\
&&\delta y(\tau)=\bar c_{21}\,e^{\lambda_1\tau}+\bar c_{22}\,e^{\lambda_2\tau},\label{perts}\eea where

\bea &&\bar c_{11}=\frac{c_{22}\delta\bar x(0)}{c_{22}c_{11}-c_{12}c_{21}},\;\bar c_{12}=-\frac{c_{12}\delta\bar y(0)}{c_{22}c_{11}-c_{12}c_{21}},\nonumber\\
&&\bar c_{21}=-\frac{c_{21}\delta\bar x(0)}{c_{22}c_{11}-c_{12}c_{21}},\;\bar c_{22}=\frac{c_{11}\delta\bar y(0)}{c_{22}c_{11}-c_{12}c_{21}}.\nonumber\eea 

As a matter of fact we do not need to compute the coefficients $\bar c_{ij}$. Actually, the structure of the eigenvalues $\lambda_i$ is the only thing we need to judge about the stability of given (hyperbolic) equilibrium points of \eqref{asode}. If the eigenvalues are complex numbers $\lambda_\pm=\nu\pm i\omega$ the perturbations \eqref{perts} do oscillations with frequency $\omega$. If the real part $\nu$ is positive the oscillations are enhanced, while if $\nu<0$ the oscillations are damped. The case $\nu=0$ is associated with a center -- harmonic oscillations -- and is not frequently encountered in cosmological applications. This latter kind of point in the phase space -- eigenvalue with vanishing real part -- is called as non-hyperbolic critical point.



\begin{thebibliography}{99}



\bibitem{stelle} K.S. Stelle, Phys. Rev. D {\bf 16} (1977) 953-969

\bibitem{stelle-1} K.S. Stelle, Gen. Rel. Grav. {\bf 9} (1978) 353-371

\bibitem{book-1} N.D. Birrell, P.C.W. Davies, ``Quantum Fields in Curved Spacetime'' (Cambridge University Press, Cambridge, 1982) 

\bibitem{book-2} I.L. Buchbinder, S.D. Odintsov, I.L. Shapiro, ``Effective Actions in Quantum Gravity'' (IOP Publishing, Bristol, 1992)

\bibitem{hindawi} A. Hindawi, B.A. Ovrut, D. Waldram, Phys. Rev. D {\bf 53} (1996) 5583-5596 [hep-th/9509142]

\bibitem{hindawi-1} A. Hindawi, B.A. Ovrut, D. Waldram, Phys. Rev. D {\bf 53} (1996) 5597-5608 [hep-th/9509147]

\bibitem{nojiri-rev-2011} S. Nojiri, S.D. Odintsov, Phys. Rept. {\bf 505} (2011) 59-144 [arXiv:1011.0544]


\bibitem{ellis-book} J. Wainwright, G.F.R. Ellis, {\it Dynamical Systems in Cosmology} (Cambridge University Press, Cambridge, 1997)

\bibitem{coley-book} A.A. Coley, {\it Dynamical Systems and Cosmology} (Dordrecht-Kluwer, Netherlands, 2003)

\bibitem{wands-prd-1998} E.J. Copeland, A.R. Liddle, D. Wands, Phys. Rev. D {\bf 57} (1998) 4686-4690 [gr-qc/9711068]

\bibitem{faraoni-grg-2013} V. Faraoni, C.S. Protheroe, Gen. Rel. Grav. {\bf 45} (2013) 103-123 [arXiv:1209.3726]

\bibitem{bohmer-rev} S. Bahamonde, C.G. Bohmer, S. Carloni, E.J. Copeland, W. Fang, N. Tamanini, Phys. Rept. {\bf 775-777} (2018) 1-122 [arXiv:1712.03107]

\bibitem{quiros-rev} I. Quiros, Int. J. Mod. Phys. D {\bf 28} (2019) 1930012 [arXiv:1901.08690]

\bibitem{quiros_ejp_rev} R. Garc\'ia-Salcedo, T. Gonzalez, F.A. Horta-Rangel, I. Quiros, D. Sanchez-Guzm\'an, Eur. J. Phys. {\bf 36} (2015) 025008 [arXiv:1501.04851]

\bibitem{quiros1} A. Avelino, R. Garc\'ia-Salcedo, T. Gonzalez, U. Nucamendi, I. Quiros, JCAP {\bf 1308} (2013) 012 [arXiv:1303.5167]

\bibitem{quiros2} R. Garc\'ia-Salcedo, T. Gonzalez, I. Quiros, M. Thompson-Montero, Phys. Rev. D {\bf 88} (2013) 043008 [arXiv:1301.6832]

\bibitem{quiros3} O. Obregon, I. Quiros, Phys. Rev. D {\bf 84} (2011) 044005 [arXiv:1011.3896]

\bibitem{quiros4} R. Garc\'ia-Salcedo, T. Gonzalez, C. Moreno, Y. Napoles, Y. Leyva, I. Quiros, JCAP {\bf 1002} (2010) 027 [arXiv:0912.5048]


\bibitem{arciniega-plb-2020} G. Arciniega, J.D. Edelstein, L.G. Jaime, Phys. Lett. B {\bf 802} (2020) 135272 [arXiv:1810.08166]

\bibitem{bueno-prd-2016} P. Bueno, P.A. Cano, Phys. Rev. D {\bf 94} (2016) 104005 [arXiv:1607.06463]

\bibitem{bueno-prd-2016-1} P. Bueno, P.A. Cano, Phys. Rev. D {\bf 94} (2016) 124051 [arXiv:1610.08019]

\bibitem{chinese} X.H. Feng, H. Huang, S.L. Li, H. Lu, H. Wei, arXiv:1807.01720

\bibitem{marciu} M. Marciu, arXiv:2003.06403

\bibitem{oliva} A. Cisterna, N. Grandi, J. Oliva, arXiv:1811.06523

\bibitem{arciniega-2} G. Arciniega, P. Bueno, P.A. Cano, J.D. Edelstein, R.A. Hennigar, L.G. Jaime, Phys. Lett. B {\bf 802} (2020) 135242 [arXiv:1812.11187]


\bibitem{hartman} P. Hartman, Proc. Amer. Math. Soc. {\bf 11} (1960) 610-620

\bibitem{grobman} D. M. Grobman, Mat. Sb. (N.S.) {\bf 56(98)} (1962) 77–94.





\end{thebibliography}
\end{document}